\documentclass{elsart}

\usepackage{epsfig}

\usepackage{amssymb}

\begin{document}

\begin{frontmatter}

\title{Empirical nonextensive laws for the county distribution of 
total personal income and gross domestic product} 

\author[ufba,cbpf]{Ernesto P. Borges}
\ead{ernesto@ufba.br}
\address[ufba]{
Escola Polit\'ecnica,
Universidade Federal da Bahia, \\
Rua Aristides Novis, 2, 40210-630 Salvador-BA, Brazil \\
}
\address[cbpf]{
Centro Brasileiro de Pesquisas F\'\i sicas, \\
Rua Dr.  Xavier Sigaud 150, %\\
22290-180 Rio de Janeiro-RJ, Brazil
}

\begin{abstract}
    We analyze the cumulative distribution of total personal income of 
    USA counties, and gross domestic product of Brazilian, 
    German and United Kingdom counties, and also of world countries. 
    We verify that generalized exponential distributions, related to
    nonextensive statistical mechanics, describe almost the whole spectrum 
    of the distributions (within acceptable errors), ranging from the 
    low region to the middle region, and, in some cases, 
    up to the power-law tail.
    The analysis over about 30 years (for USA and Brazil) shows a regular 
    pattern of the parameters appearing in the present phenomenological 
    approach, suggesting a possible connection between the underlying dynamics 
    of (at least some aspects of) the economy of a country 
    (or of the whole world) and nonextensive statistical mechanics.
    We also introduce two additional examples related to geographical
    distributions: land areas of counties and land prices, and the
    same kind of equations adjust the data in the whole range of the spectrum.
\end{abstract}

\end{frontmatter}

\section{Introduction}

Despite of the ubiquity of Gaussians in nature, there are many,
also ubiquitous, examples of non-Gaussian distributions.
Power-laws, for instance, appear in a variety of physical, biological, 
psychological, and social/economical phenomena 
\cite{pareto,zipf,mandelbrot,Stanley:Nature2001,hydra,urban,bak:earthquakes,%
abe:earthquakes,internet}.
Frequently, such systems do not exhibit power-law behavior in the entire
spectrum, but rather power-law {\em tails}.
Characterization of economical systems, more specifically the distribution of
personal income, are usually assumed to follow Pareto's law \cite{pareto},
\begin{math}
 p(x) \propto x^{-1-\alpha},
\end{math}
in the large income region (typically $1\ge \alpha \ge 2$), 
and a log-normal distribution \cite{gibrat},
\begin{eqnarray}
 p(x)=\frac{1}{x\sqrt{2\pi \sigma^2}}
 \exp\left[-\frac{\log^2(x/x_0)}{2\sigma^2}\right],
\end{eqnarray}
in the middle (or low-middle) income region ($p(x)$ is the probability 
density function, $x_0$ is a mean value and $\sigma^2$ is a variance).
See \cite{souma2000,souma2002,ishikawa} for recent revisiting of this
approach to the problem.

It is already known the existence of connections between economical systems 
(financial markets) and nonextensive statistical mechanics \cite{ramos-2000} 
(see \cite{ct-lisa:2003} for a recent review). 
In the present work we address a different feature of economical systems:
the distribution of total personal income (PI) of counties,
as well as total gross domestic product (GDP) of counties for a given country
(both PI and GDP can be an index for the value added).
We similarly consider distribution of GDP of the {\em countries} of the world.
We use distributions that belong to the family of the $q$-exponential function
\cite{ct:quimicanova,epb:jpa1998},
\begin{eqnarray}
 \label{eq:qexp}
 \exp_q (x) \equiv [1 + (1-q) x]_+^{\frac{1}{1-q}}
\end{eqnarray}
($q\in \mathbb{R}$, $[\cdots]_+ \equiv \mbox{max}\{\cdots,0\}$), 
that naturally emerge from nonextensive statistical mechanics 
\cite{CT1988,Curado-CT1991,CT-Mendes-Plastino}
(for recent reviews and updated bibliography, see 
Ref.~\cite{abe-springer,cagliari,http}).
$q$-Exponentials (with negative argument, which is the case we are
interested in here; hereafter we will consider $\exp_q(-x)$, with $q\ge1$ 
and $x>0$) 
present asymptotic power-law tails, as many complex systems do.
Along these lines, we are able to describe (almost) the whole spectrum 
of the distribution (and not only the tails) with a single function, 
which points towards an unified approach of the problem.
In a certain sense, this problem resembles another one, namely the number 
of citations of scientific papers, which likewise present power-law behavior 
only at the tail.
It was first conjectured that different phenomena rule large-cited and
low-cited papers (see Ref.~\cite{redner} and references therein). 
A nonextensive approach to the problem \cite{ct-marcio} showed that it is 
possible to have a single function describing the whole spectrum of citations.

\section{Distribution generators}

Let us formulate the problem by two alternative paths.
One way of characterizing a distribution is through the variational approach,
in which an {\em entropy} is maximized under constraints of normalizability
and finiteness of a generalized moment of the distribution
($\langle |x|^{\gamma} \rangle = \mbox{constant}$, $\gamma > 0$)
(see, e.g., Ref.~\cite{balian}).
For instance, if we take into account Boltzmann-Gibbs entropy,
\begin{eqnarray}
 S = - k_B\int  p(x) \ln p(x) \, dx,
\end{eqnarray}
submitted to the constraints of normalizability
\begin{eqnarray}
 \label{eq:norma}
 \int p(x)\,dx=1
\end{eqnarray}
and finiteness of a certain momentum of order $\gamma$ 
\begin{eqnarray}
 \label{eq:momentum}
 \int |x|^\gamma\,p(x)\,dx < \infty,
\end{eqnarray}
it comes out exponential forms (stretched exponentials),
\begin{eqnarray}
 \label{eq:exponential}
 p(x)\propto \exp (-\beta x^{\gamma}), 
\end{eqnarray}
with $\beta$ being the Lagrange multiplier.
Typically $\gamma=1$ when we are dealing with distributions of, e.g., energy, 
and $\gamma=2$ for distributions of space positions, i.e., diffusion.
For the sake of generality, we put not only integer values of
$\gamma$, but rather $\gamma\in\mathbb{R}$; that's why it appears the modulus
in Eq.~(\ref{eq:momentum}) (although in the forthcoming examples we only 
consider $\gamma=1$ or $\gamma=2$, and also $x>0$, which makes unnecessary
the modulus).
$\gamma=1$ yields exponentials, $\gamma=2$, Gaussians.

If, instead, we take nonextensive entropy \cite{CT1988}
\begin{eqnarray}
 S_q \equiv k\frac{1-\int [p(x)]^q \, dx}{q-1} \qquad 
 (q\in \mathbb{R}),
\end{eqnarray}
($k$ is a non-negative constant, related and possibly equal to Boltzmann's
constant $k_B$), with the same normalizability constraint, Eq.~(\ref{eq:norma}),
and a $q$-generalized version of the finiteness of the momentum of order
$\gamma$,
\begin{eqnarray}
 \label{eq:q-momentum}
 \int |x|^\gamma\,[p(x)]^q\,dx < \infty,
\end{eqnarray}
then $q$-stretched exponentials appear:
\begin{eqnarray}
 \label{eq:q-exponential}
 p(x)&\propto&\exp_q (-\beta_q\,x^{\gamma}) \nonumber \\
     &\propto &[1-(1-q)\beta_q\,x^{\gamma}]_+^{\frac{1}{1-q}}.
\end{eqnarray}
$q=1$ recovers usual stretched exponentials, with $\beta_q\equiv\beta$,
Eq.~(\ref{eq:exponential}).
$q\ne1$ and $\gamma=1$ recovers the $q$-exponential itself, Eq.~(\ref{eq:qexp}).
Functions with $\gamma=2$ and $q\ne1$ can consistently be called $q$-Gaussians.
Particular cases are, of course, the Gaussian distribution ($q=1$), 
and the Lorentzian distribution ($q=2$).
This path was followed by Ref. \cite{CT-Levy,CT-Prato}.

\bigskip

An alternative way of characterizing a distribution is through the
{\em differential equation} it satisfies. 
This path was developed in Ref. \cite{Tsallis-Bemski-Mendes}, 
within the framework of nonextensive statistical mechanics, 
but it was originally formulated in Planck's celebrated first papers 
on black-body radiation law \cite{planck}, the birth of quantum mechanics.
See Ref.~\cite{ct:physicaD-2003} for a comprehensive review on this
differential equation approach, and also for historical remarks about 
Planck's first papers on black-body radiation.
Stretched exponential distributions obey 
\begin{eqnarray}
 \frac{1}{\gamma x^{\gamma-1}} \frac{dp}{dx} = - \beta\,p,
\end{eqnarray}
while nonextensive $q$-stretched exponentials follow a simple generalization
of the former equation:
\begin{eqnarray}
 \label{diff.eq:qexp}
 \frac{1}{\gamma x^{\gamma-1}} \frac{dp}{dx} = - \beta_q\,p^q,
\end{eqnarray}
whose solution is given by Eq.~(\ref{eq:q-exponential}).

Some complex systems, e.g. re-association of oxygen in folded myoglobin 
\cite{Tsallis-Bemski-Mendes},
linguistics \cite{montemurro},
cosmic rays \cite{cronin,cosmic:pla,cosmic:creta},
economical systems (distribution of returns of New York Stock Exchange
\cite{Plerou}),
and also the economical examples we are dealing here,
exhibit not one, but {\em two} power-law regimes (i.e., two different slopes
in a log-log plot, according to the value of the independent variable), 
with an usually marked crossover between them, sometimes referred to as the 
{\em knee}. Such behavior demands a probability distribution obeying a more 
general differential equation than the two former ones, namely
\begin{eqnarray}
 \label{eq:q,q'-gaussian}
 \frac{1}{\gamma x^{\gamma-1}} \frac{dp}{dx} =
 - (\beta_q-\beta_{q'})\,p^q - \beta_{q'}\,p^{q'}
\end{eqnarray}
($1\le q' \le q$, $0 \le \beta_{q'} \le \beta_q$),
where $q$ and $q'$ are connected to the two different slopes of the regimes
(the asymptotic slope in a $\log$-$\log$ plot is given by $\gamma/(1-q)$).
$q$ controls the slope of the first (intermediate) power-law regime, 
and $q'$, the second (the tail).
The solution of Eq.~(\ref{eq:q,q'-gaussian}) is expressed in terms of
hypergeometrical functions (see Ref.~\cite{Tsallis-Bemski-Mendes}
for the analytical expression), and  we can consistently call such functions
$(q,q')$-stretched exponentials.
They naturally present a crossover (knee) at
\begin{eqnarray}
 \label{eq:knee}
  x_{knee}^{\gamma} = \frac{[(q-1) \beta_q]^{\frac{q'-1}{q-q'}}}
  {[(q'-1)\beta_{q'}]^{\frac{q-1}{q-q'}}}.
\end{eqnarray}
Notice that $\beta_{q'} = 0$, or $\beta_{q'} = \beta_q$, or even
$q' = q$ recover the differential equation obeyed by $q$-stretched exponentials,
Eq.~(\ref{diff.eq:qexp}), with $x_{knee} \to \infty$.

Another particular case of Eq.~(\ref{eq:q,q'-gaussian}) is obtained with 
$q'=1$. 
It is curious to note that this differential equation (with $q'=1$) 
is known as Bernoulli's equation \cite{BoyceDiPrima,Kreyszig}, 
and it is exactly the one used by Planck in his October 1900 paper 
\cite{planck} (with $q=2$, $q'=1$, $\gamma=1$)%
\footnote{It is also worth mention that Planck adopted this equation as a 
fitting procedure (and certainly with a great amount of physical intuition). 
In his words \protect\cite{planck}, 
``one gets a radiation formula with two constants \dots which, as far as 
I can see at the moment, fits the observational data, published up to now, 
as satisfactorily as the best equations put forward for the spectrum \dots''}.
The distribution of this $q'=1$ case is given by
\cite{Tsallis-Bemski-Mendes}
\begin{eqnarray}
 p(x)=\left[ 1-\frac{\beta_q}{\beta_1}
 +\frac{\beta_q}{\beta_1} e^{(q-1)\beta_1 x^\gamma} \right]^{\frac{1}{1-q}},
\end{eqnarray}
and presents an intermediate power-law regime, followed by an {\em exponential}
tail, with a crossover at
\begin{eqnarray}
 x_{knee}^{\gamma}=\frac{1}{[(q-1)\beta_1]}.
\end{eqnarray}
An example of system that may be described by this ($q\ne 1,\;q'=1$) case is
the measure of success of musicians \cite{epb:musicians}.

\section{Geographical distribution of Total Personal Income and 
Gross Domestic Product}

We consider Eq.~(\ref{eq:q,q'-gaussian}) with 
$p\equiv P$ being the {\em inverse cumulative probability distribution},
$P(X\ge x)=\int_x^\infty dy\;p(y)$
($P(X\ge x)$ is the probability of finding the distribution variable with
a value $X$ equal to, or greater than, $x$),
and $x\equiv x/x_0$ is the ratio between an economical variable and its
minimum value:
in the discrete case, $x_i \equiv x_i/x_{min}$, where $x$ stands for 
the economical variable, in our analysis, PI of a county, or GDP of a county 
(or of a country). Index $i$ refers to the county (or country), 
and $min$ is the poorest (lowest ranking) county (country). 

We analyze one case of PI county distribution, USA counties
(for years ranging from 1970 to 2000) 
\cite{source:usa}, 
and three cases of GDP county distribution:
Brazilian counties (from 1970 to 1996) 
\cite{source:brasil}, 
German counties (from 1992 to 1998) 
\cite{source:germany}, 
and United Kingdom counties (from 1993 to 1998) 
\cite{source:uk}.
(See \cite{source:usa} for the method of calculation of USA county PI.)
All these cases are well described with $\gamma=2$, i.e., $(q,q')$-Gaussians.

\begin{figure}[htb]
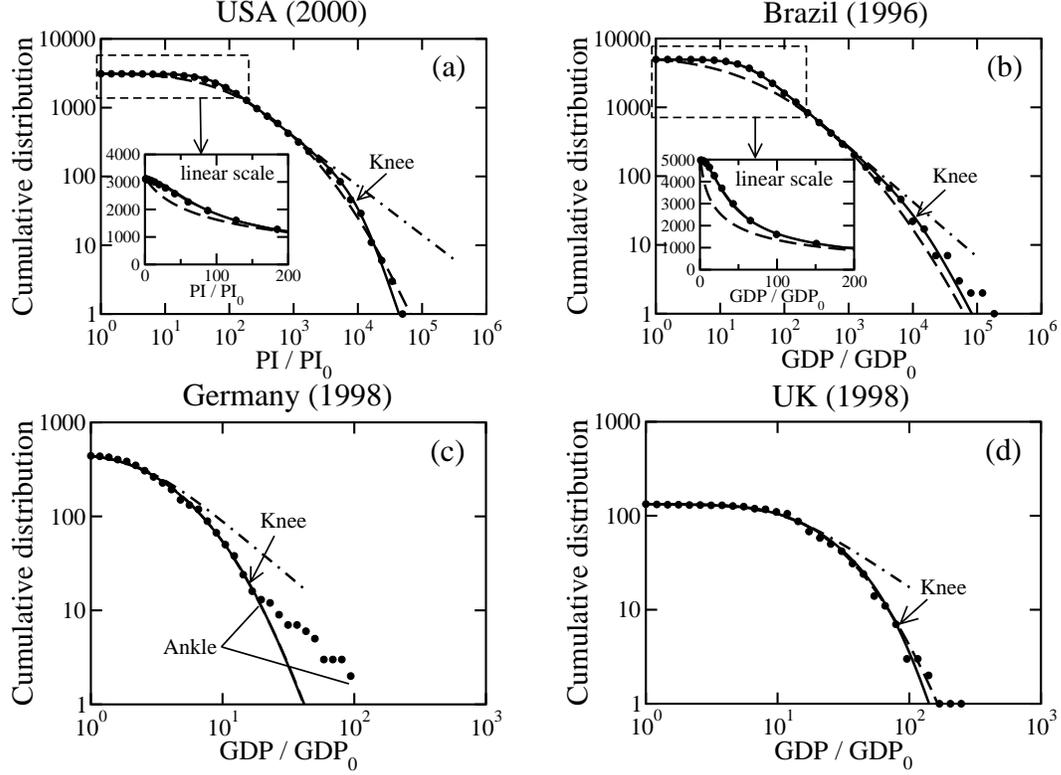

 \begin{minipage}[htb]{0.47\textwidth}
  \epsfig{figure=usa2000-ratio.eps,width=\textwidth,clip=}
 \end{minipage}
 \hfill
 \begin{minipage}[htb]{0.47\textwidth}
  \epsfig{figure=br1996-ratio.eps,width=\textwidth,clip=}
 \end{minipage}

 \begin{minipage}[htb]{0.47\textwidth}
  \epsfig{figure=germany1998-ratio.eps,width=\textwidth,clip=}
 \end{minipage}
 \hfill
 \begin{minipage}[htb]{0.47\textwidth}
  \epsfig{figure=uk1998-ratio.eps,width=\textwidth,clip=}
 \end{minipage}
 \caption{
  Binned inverse cumulative distribution of county PI/PI$_0$ (USA) and 
  GDP/GDP$_0$ (Brazil, Germany and UK). Three distributions are displayed
  for comparison: (i) $q$-Gaussian (with $\beta_{q'}=0$) (dot-dashed),
  (ii) $(q,q')$-Gaussian (solid), and (iii) log-normal (dashed lines).
  Figures (a) and (b) present Insets with linear-linear scale, to make more 
  evident the quality of the fitting at the low region 
  (In Fig.s (c) and (d), the ($q$,$q'$)-Gaussian and the log-normal curves 
  are superposed and so are visually indistinguishable).
  The positions of the knees (according to Eq.~(\protect\ref{eq:knee})) 
  are indicated.
  The ankle is particularly pronounced in (c), though it is 
  also present in the other cases.
 }
 \label{fig:distributions}
\end{figure}

Fig.~\ref{fig:distributions} illustrates the results with inverse cumulative 
distributions. Inverse cumulative distribution, or the rank, is equal to
the number of counties $N_{counties}$ times $P$, with $P$ given by the 
corresponding cumulative distribution probability.
Three curves are shown in each Fig.~\ref{fig:distributions}(a)--(d): 
(i) $q$-Gaussian distributions, which can describe low range data,
(ii) $(q,q')$-Gaussian, which shows to be able to reproduce the low-middle 
range, including the knee,
and (iii) log-normal distributions, that were adjusted to fit middle 
range values. 
For USA and Brazil, we observe that the $(q,q')$-Gaussian describes 
the data in almost the entire range; 
for Germany and UK, both $(q,q')$-Gaussian and log-normal are able to describe 
the data in the low-middle region (the curves are practically visually 
indistinguishable in this region). For USA and Brazil, 
the log-normal distribution fails in the low region --- see Inset of 
Fig.s~\ref{fig:distributions}(a) and \ref{fig:distributions}(b).
Values of the parameters are given in Table~\ref{tab:parameters}.

At a first glance, it might seem that log-normal distributions are
more parsimonious (and consequently, preferable) in the description of 
these problems than the $(q,q')$-Gaussians, once the former has two fitting 
parameters, while the later has four parameters.
But when we look in detail to the problem, we realize that in many
cases the log-normal is able to describe just the middle range values 
of the distributions (sometimes low-middle range). Deciding where this 
middle range begins and where it ends works as if there were two additional 
hidden parameters in this log-normal distribution. When this happens, both 
log-normal and $(q,q')$-Gaussians present the same fitting degree of freedom.

\begin{table}[htb]
\caption{Parameters for the distribution functions, for the years shown in
Fig.~\protect\ref{fig:distributions}.}
\label{tab:parameters}
\begin{tabular}{lcccccccc}
\hline
Country & Year & $N_{counties}$ & $q$ & $q'$ & $1/\sqrt{\beta_{q}}$ &
$1/\sqrt{\beta_{q'}}$ & $x_0$ & $\sigma$ \\ \hline
USA     & 2000 & 3110 & 3.80  & 1.7 & 87.71 & 2236.07 & 110  &  7   \\
Brazil  & 1996 & 4973 & 3.50  & 2.1 & 40.82 &  816.50 &  22  & 10   \\
Germany & 1998 &  440 & 2.70  & 1.5 &  3.16 &    6.59 &  3.5 &  1.5 \\
UK      & 1998 &  133 & 3.12  & 1.4 & 18.26 &   37.80 &  20  &  1.5 \\ 
\hline
\end{tabular}
\end{table}

Large GDP range displays a different behavior:
the distribution presents a {\em second} crossover,
bending upwards and giving rise to a different (third) power-law regime.
This effect is very pronounced for Germany, and in a smaller degree for UK, 
while for USA and Brazil, it is almost hidden in the binned distribution 
(as shown in Fig.~\ref{fig:distributions}), but it is visible with 
unbinned plots%
\footnote{In a binned distribution, the ordinate shows the number of data 
(normalized or not) that falls within a (usually small) region, or bin, 
in the abscissa. 
In the distributions shown in Fig.~\protect\ref{fig:distributions}, 
the bins are logarithmic equally spaced, i.e., their width are exponentially 
increasing. 
In unbinned distributions, each point in the figure corresponds to an 
original data. Binned distribution was chosen in 
Fig.~\protect\ref{fig:distributions} for better visualization.}.
In USA, for instance, only the two major counties (Los Angeles and Cook 
(part of Chicago)) belong to this regime. 
Similarly, in Brazil, we have S\~ao Paulo and Rio de Janeiro within this
regime.
This feature is commonly exhibited by various systems, sometimes 
referred to as {\em king} effect \cite{Laherrere1998}. 
It is also present in highly energetic cosmic rays, been referred 
to as {\em ankle} \cite{cronin} (we adopt this nomenclature in the Figures). 
Such behavior is possibly related to nonequilibrium phenomena, or 
(at least in some cases) poor statistics, and lies outside the 
present approach. We recall that the number of counties in USA and Brazil
is about one order of magnitude greater than that of Germany and UK, and
this possibly is related to the more pronounced king effect in these last
to countries.

\begin{figure}[htb]
 \begin{center}
 \epsfig{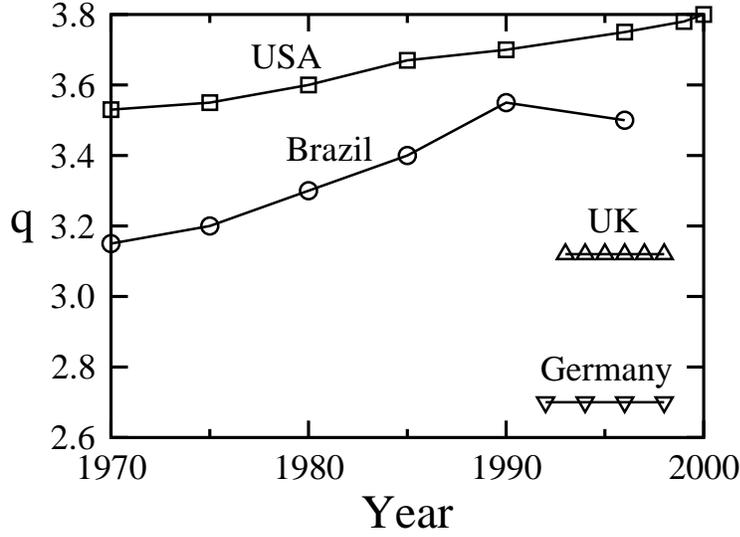}
 \caption{
  Evolution of parameter $q$ for USA (squares), Brazil (circles),
  UK (up triangles) and Germany (down triangles).
  The parameters $q'$ (for each country) are constant for all years:
  $q'_{Brazil}=2.1$, $q'_{USA}=1.7$, $q'_{Germany}=1.5$, $q'_{UK}=1.4$.
  Lines are only guide to the eyes.
 }
 \label{fig:q}
 \end{center}
\end{figure}

Fig.~\ref{fig:q} shows temporal evolution of the parameter $q$.
USA present an approximately uniform increase of $q$
over 30 years.
In the case of Brazil, the tendency of increase from 1970 to 1990
was broken from 1990 to 1996.
Germany and UK present constant values of $q$ over the years
for which there are available data.
The increase of $q$ (observed for USA and Brazil) indicates
increasing {\em inequality}: the greater the $q$, long-lasting the tail,
the greater the probability of finding counties much richer than others.
The parameter $q'$ (for a given country) is taken constant for all years.
The smaller values of $q$ and $q'$ for Germany and UK,
when compared to USA and Brazil, reflect the well balanced distribution
of value added of these European countries, relative to the analyzed American
countries.
The relation between the slopes (related to $q$) and equality/inequality
is not a new conclusion; it is known since Pareto \cite{pareto} (see also 
Ref.~\cite{solomon} and references therein).

\begin{figure}[htb]
 \begin{center}
 \epsfig{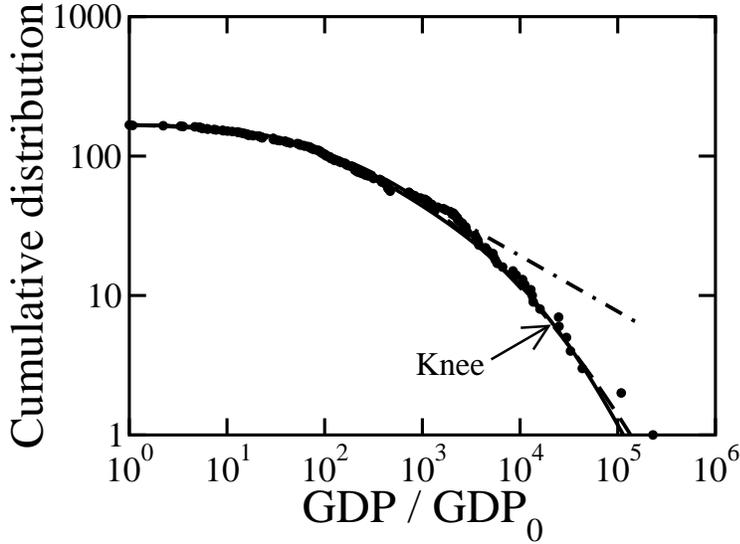}
 \caption{
  Inverse cumulative distribution of GDP/GDP$_0$ of 167 {\em countries} 
  for the year 2000 (unbinned data: each point corresponds to a country). 
  The data are fitted with $(q,q')$-exponential (solid) and log-normal 
  (dashed line) distributions --- they are visually indistinguishable for 
  this example. 
  $q$-exponential (with $\beta_{q'}=0$, dot-dashed) is also shown for 
  comparison. Values of the parameters are $q=3.5$, $q'=1.7$, 
  $1/\beta_q=111.1$, $1/\beta_{q'}=2500.0$.
  The knee, according to Eq.~(\protect\ref{eq:knee}), is located at 
  GDP/GDP$_0=19\,665$. Log-normal curve with $x_0=220$ and $\sigma=13$.
 }
 \label{fig:world}
 \end{center}
\end{figure}

We have also analyzed the world GDP {\em country} distribution, for the
year 2000 \cite{source:world}. In this case, we found that a 
$(q,q')$-exponential ($\gamma=1$) fits better the data than a 
$(q,q')$-Gaussian ($\gamma=2$) in the low-middle region.
Although the difference between the two functions (with $\gamma=1$ and 
$\gamma=2$) is perceptible, it is not that much big. This purely 
phenomenological observation deserves further investigation to corroborate 
or not our results. 
If it is confirmed to be $\gamma=1$, a possible interpretation might be due to 
the nature of interactions between countries, which is expected to differ from 
the interactions between counties inside a country.
Fig.~\ref{fig:world} shows the results. The king effect is also present
here, particularly for the two major GDP countries, USA and Japan.

\section{Distribution of land areas and land markets}

In this section we add two different examples, related to geographical 
distributions: (i) distribution of land areas of Brazilian counties, and 
(ii) distribution of Japan land prices.

\begin{figure}[htb]
 \begin{center}
 \epsfig{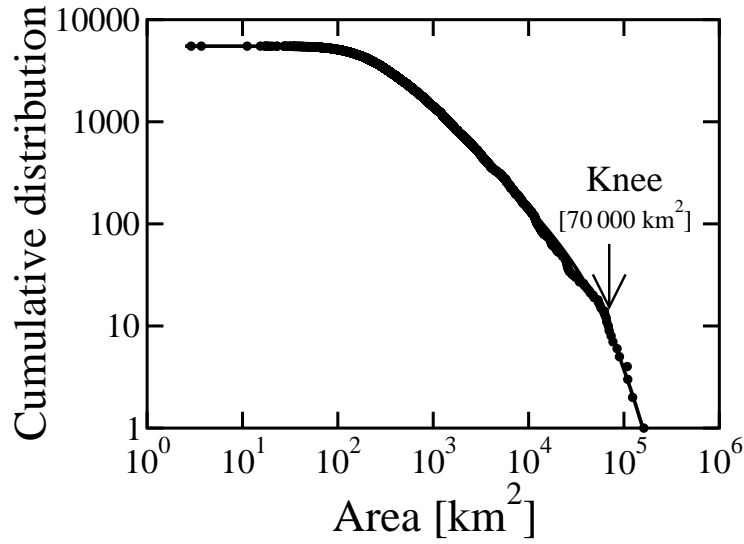}
 \caption{
  Inverse cumulative distribution of land areas of Brazilian counties 
  (unbinned data). 
  Solid line is a $(q,q')$-Gaussian. $q=3.07$, $q'=1.56$, 
  $1/\sqrt{\beta_q}=353.55$ km$^2$, $1/\sqrt{\beta_{q'}}=11\,226.7$ km$^2$.
 }
 \label{fig:brazilian-areas}
 \end{center}
\end{figure}

Let us focus on the first example, illustrated by 
Fig.~\ref{fig:brazilian-areas}.
The minor Brazilian county has 2.9 km$^2$ (Santa Cruz de Minas, in the State 
of Minas Gerais), and the major one has $161\,446$ km$^2$ (Altamira, in the 
State of Par\'a, within the Amazon forest%
%, greatest than countries like Austria, Portugal, Cuba; it is about the area 
%of Tunisia%
)
\cite{ibge}.
There are many causes for a county to have a given area, including, 
among others, geographical, political, demographical and economical factors.
The $(q,q')$-Gaussian fits (within an acceptable error) practically all 
county areas (more than 5500, in the year 1998), from the smaller up to 
the greater.

\begin{figure}[htb]
 \begin{center}
 \epsfig{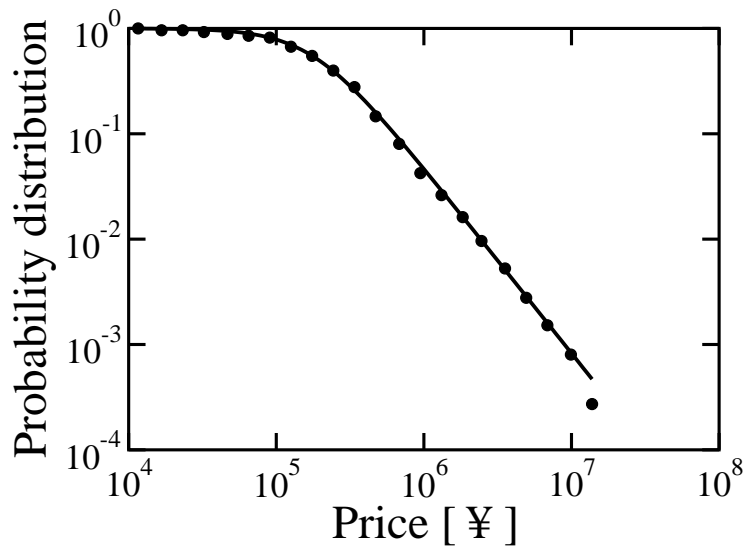}
 \caption{
  Inverse cumulative probability distribution of Japan land prices 
  for the year 1998.
  The data (binned) were taken from Fig.~1 of \protect\cite{kaizoji}. 
  %Total of points is 30,600. 
  Solid curve is a $q$-Gaussian with $q=2.136$, which
  corresponds to the slope $-1.76$ (found by the Author of 
  \protect\cite{kaizoji}), and $1/\sqrt{\beta_q}=188\,982$ Yen.
 }
 \label{fig:japanlandprices}
 \end{center}
\end{figure}

Now let us consider the problem of Japan land prices, recently addressed
\cite{kaizoji}. The author found a power-law tail for the cumulative 
probability distribution of land price, with a slope of $-1.7$
($P(X\ge x) \propto x^{-1.7}$). Fig.~\ref{fig:japanlandprices} makes evident 
that the $q$-Gaussian (with $\beta_{q'}=0$) fits the whole range of data 
(except the point with the higher price) and not only the tail 
(we recall that the probability distribution, from \cite{kaizoji}, is binned 
--- maybe with unbinned data, we could find $\beta_{q'}\ne0$, and with this, 
it would possibly include the last point(s) under the curve).

\section{Final remarks}

Finally we would like to point out that $q$-Gaussians and log-normal 
distributions seem to be equally able to describe the data in the low-middle 
region. As these data usually range not so many decades, one cannot 
unequivocally decide the `correct' distribution by simple comparison, 
and such phenomenological approaches (ours included) only give hints 
about the underlying dynamics of the economy.
$q$-Gaussians present power-law tails (log-normal distributions don't), 
which is an important characteristic when dealing with complex systems.
The good concordance of the present procedure with data,
including the smoothness of temporal evolution of the parameters,
together with previous works along these lines \cite{ramos-2000,ct-lisa:2003},
may suggest a new path for investigating economical relations,
namely the development of models based on the framework of
nonextensive statistical mechanics.
Such $q$-distributions (or $(q,q')$-distributions) appear when 
long-range interactions, long-term memory, (multi)fractality and/or 
small-world networking are present, expected features in complex systems 
(including economical ones).
This further central step (development of models that essentially describe
the underlying microscopic dynamics of the systems) is certainly not an
easy task. The answer (as suggested in Ref.~\cite{ct:physicaD-2003}) 
may come from Barab\'asi-Albert's approach to the problem of small-world 
networks \cite{barabasi-albert:science-1999,albert-barabasi:prl-2000}, 
considering preferential attachment of new vertices, that are added to a 
network, to sites that are already well connected, leading to scale-free 
power-law distributions.

These examples join many others (some cited in the Introduction) that are 
fairly fitted by equations related to nonextensive statistical mechanics. 
Of course these are only empirical observations, but there is a significant
amount of evidence that nonextensivity and complex systems are intimately 
connected 
(we don't mean that {\em all} complex systems might be somehow related to
nonextensivity, but it is possible that such complex systems may be divided
into classes of universality, some of them presenting a nonextensive nature).
Besides, it is already known how to estimate {\em a priori} the $q$ index for
some classes of systems, namely low-dimensional dissipative maps, 
based on its dynamics (see \cite{fulvio-alberto:1,fulvio-alberto:2} 
and references therein).
Of course these are much more simple systems than the ones we are dealing with,
but many efforts have being made to extend this predictive nature to other 
more complex systems (see also \cite{ct:physicaD-2003}). 
These results point towards the interpretation that the
nonextensive indexes $q$ and $q'$ are not merely  fitting parameters, 
but they are related to more fundamental dynamic features of the problem. 
With this point of view in mind, the present proposed approach, 
at the moment being just a phenomenological observation, 
may happen to be of a more fundamental nature, 
and not simply a fitting procedure.

\section*{Acknowledgments}

Stimulating discussions with C. Tsallis, L.G. Moyano, 
G.F.J. A\~na\~nos and F. Baldovin are greatly acknowledged.
We also acknowledge the referees, whose remarks improved the paper.
This work was partially supported by CAPES (Brazilian agency).

\newpage

\begin{center}
{\Large Figures}
\end{center}
\vspace{2cm}

\end{document}